\documentclass[twocolumn,pra,showpacs,superscriptaddress]{revtex4-1}
\usepackage{amsmath} %
\usepackage{amssymb} %
\usepackage{graphicx} %
\usepackage{bm} %
\usepackage{epsfig} %
\usepackage{mathrsfs} %
\usepackage{physics} %
\usepackage{subfig} %

\begin{document}

\title{Scattering of a diatomic composite system}

\author{Tieling Song}

\affiliation{Institute of Physics, Beijing National Laboratory for
  Condensed Matter Physics, Chinese Academy of Sciences, Beijing
  100190, China}

\author{Wei Zhu}

\affiliation{Institute of Physics, Beijing National Laboratory for
  Condensed Matter Physics, Chinese Academy of Sciences, Beijing
  100190, China}

\author{D.~L. Zhou}

\email{zhoudl72@iphy.ac.cn}

\affiliation{Institute of Physics, Beijing National Laboratory for
  Condensed Matter Physics, Chinese Academy of Sciences, Beijing
  100190, China}

\affiliation{School of Physical Sciences, University of Chinese
  Academy of Sciences, Beijing 100190, China}

\date{\today}

\begin{abstract}
  We investigate the scattering problem of a two-particle composite
  system on a delta-function potential. Using the time independent
  scattering theory, we study how the transmission/reflection
  coefficients change with the height of external potential, the
  incident momentum, and the strength of internal potential. In
  particular, we show that the existence of internal degree of freedom can
  significantly change the transmission/reflection coefficients even
  without internal excitation. We consider two scenarios: the internal degree
  of freedom of the incident wave is set to be in a state with even parity or
  odd parity, we find that the influence of a symmetric Hamiltonian
  is greater on the odd-parity internal states than on the even-parity ones.
\end{abstract}



\maketitle

\section{Introduction}

The quantum superposition principle lies at the heart of quantum
mechanics and allows massive objects to be prepared in spatial
superposition of the order of their sizes. Although quantum
interferences of macroscopic objects have remaind experimentally
challenging~\cite{ORom,ABas}, a series of preparatory work
~\cite{WMar,ORome,ORomer,TKov,UBHo,JQLi,MCar,MAbd} have been finished.
Experiments involving composite systems~\cite{WSch,MArn,SGer} provide
a way to probe the quantum interferences of macroscopic objects,
moreover, it requires controlled splitting of a wave pocket to observe
interference. In our model, we simulate this process by employing a
delta-function potential to separate the incoming wave into two
(reflected and transmitted) components. As a composite system that
contains two particles at least, the energy levels of the internal
degree of freedom may be excited. In the following, we consider a
diatomic bound system and exhibit how the internal states affect the
reflected and transmitted components.

As a good approximation to many actual phenomena, quantum mechanical
scattering in one dimension attracts increasing interest during the
past years~\cite{LLSa,AMos1,WTrz,LVChe,MGRo,LVCh,FQue}. Scattering
theory~\cite{JRTa,RGNe} promotes greatly the experimental research on
the interaction and internal structure of particles. The elegance and
power of the $S$-matrix formulation is beyond doubt, but this
formulation always has high computation complexity, especially for the
higher-order correction. In this paper, we propose a simple method to
calculate the probability that a composite system that entered the
collision with in asymptote state will be observed to emerge with out
asymptote state. $S$-matrix within Born approximate is also calculated
to compare with our proposal.

In our study, we discuss the scattering process of a two-particle bound system to
mimic the splitting of the incoming wave corresponding to a diatomic composite system.
The coefficients of reflection and transmission for different internal modes are
worked out by invoking appropriate boundary conditions on the eigenfunctions of the
Hamiltonian, but not by calculating the high-order correction of $S$-matrix elements.
In the following, we will see the condition that excited internal modes become populated
and their influence on the reflected and transmitted components. We
propose our model in section II and list the results in section III,
in section IV, we end with a summary.

\section{Theoretical Model}

Consider two particles with mass $m_{1},m_{2}$ and coordinates
$x_{1},x_{2}$. A model which resembles the interaction of these
particles and the scattering potential is specified by the Hamiltonian
\begin{equation}
  H = -\frac{\hbar^{2}}{2m_{1}} \pdv[2]{x_{1}}  -
  \frac{\hbar^{2}}{2m_{2}} \pdv[2]{x_{2}} + \Omega^{2}
  {(x_{1}-x_{2})}^{2} + \gamma_{1} \delta(x_{1}) +  \gamma_{2} \delta(x_{2}),
  \label{H}
\end{equation}
where we assume that the particles are tied to each other by a
harmonic coupling with stiffness $\Omega$ and the scattering potential
has the form $V(x_{i})=\gamma_{i}\delta(x_{i})$.

For convenience, we rewrite the Hamiltonian in terms of
center-of-mass coordinate $X=(m_{1}x_{1}+m_{2}x_{2})/(m_{1}+m_{2})$
and relative coordinate $x=x_{2}-x_{1}$
\begin{equation}
  H = -\frac{\hbar^{2}}{2M} \pdv[2]{X} - \frac{\hbar^{2}}{2\mu}
  \pdv[2]{x} + \frac{1}{2} \mu \omega^{2} x^{2} + \gamma_{1}
  \delta(X-r_{2}x) +  \gamma_{2} \delta(X+r_{1}x),
  \label{H1}
\end{equation}
where $M=m_{1}+m_{2}$, $\mu=m_{1}m_{2}/M$, $r_{i}=m_{i}/M$.

\begin{figure}[htbp]
  \centering
  \includegraphics[width=10cm]{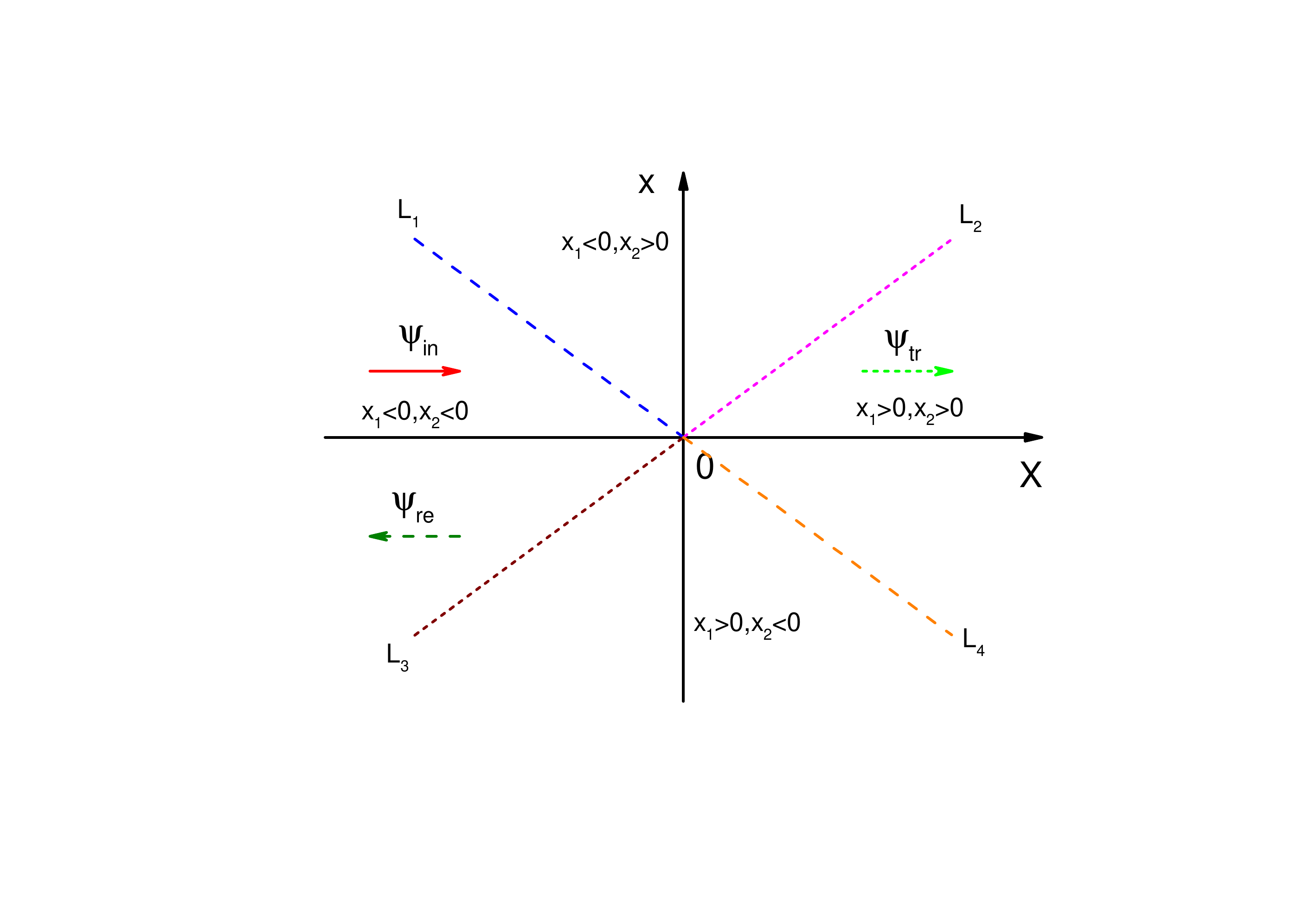}
  \caption{(Color online) \textbf{Schematic scattering process in the
      $X-x$ frame}. The scattering potentials locate along $x_{1}=0$
    (i.e., $L_{2}$ and $L_{3}$) and $x_{2}=0$ (i.e., $L_{1}$ and
    $L_{4}$), and an incoming wave is separated into a reflected and a
    transmitted component.}~\label{schematic}
\end{figure}

Obviously, the eigenstates of
$H_{0}=-\hbar^{2}\pdv[2]{X}/2M-\hbar^{2}\pdv[2]{x}/2\mu+\mu\omega^{2}x^{2}/2$
are
\begin{equation}
  \phi_{K,n}(X,x) = \frac{1}{\sqrt{2\pi}} e^{iKX} \psi_{n}(x),
  \label{eigenstateh0}
\end{equation}
for $n=0,1,2...$ with eigenenergies
\begin{equation}
  E_{K,n} = \frac{ \hbar^{2} K^{2}}{2M} + \left( n + \frac{1}{2}
  \right) \hbar \omega,
  \label{eigenenergyh0}
\end{equation}
where $K$ denotes the momentum of the center-of-mass and
$\{\psi_{n}(x)\}$ are the normalized stationary wave functions for
harmonic oscillator and they describe the internal states of this
bound system. If the system is incoming from the left (as
shown in Fig.~\ref{schematic}) with momentum $K_{0}$ and the internal
degree of freedom is assumed to be in the $l$-th state,  the
scattering state reads
\begin{align}
  \Psi_{K_{0},l}^{+}
  & = &
        \begin{cases}
          \phi_{K_{0},l} + \sum_{n} \alpha_{n} \phi_{-K_{n},n}, &
          x_{1}<0,x_{2}<0\\
          \sum_{n} \mu_{n} \phi_{K_{n},n} + \sum_{n} \nu_{n}
          \phi_{-K_{n},n}, & x_{1}<0,x_{2}>0\\
          \sum_{n} \xi_{n} \phi_{K_{n},n} + \sum_{n} \eta_{n}
          \phi_{-K_{n},n}, & x_{1}>0,x_{2}<0\\
          \sum_{n} \beta_{n} \phi_{K_{n},n}, & x_{1}>0,x_{2}>0
        \end{cases}
\end{align}
where $\{\alpha_{n},\beta_{n}\}$ are the amplitudes of different modes for
the reflected and transmitted waves respectively, $K_{n}$ are the momentum of the center-of-mass
corresponding to the $n$-th internal state and can be derived from the energy conservation condition:
\begin{equation}
  \frac{\hbar^{2} K_{0}^{2}}{2M} + (l + \frac{1}{2}) \hbar \omega =
  \frac{\hbar^{2}K_{n}^{2}}{2M} + (n+\frac{1}{2}) \hbar \omega.
  \label{energyconser}
\end{equation}
Furthermore, we can
easily get the probability current densities of incident, reflected and
transmitted waves with different modes
\begin{align}
  J_{0}^{in} &= \frac{1}{2 \pi} \frac{\hbar K_{0}}{M}, \label{J0in}\\
  J_{n}^{re} &= - \frac{|\alpha_{n}|^{2}}{2 \pi} \frac{\hbar K_{n}}{M}, \label{Jnre}\\
  J_{n}^{tr} &= \frac{|\beta_{n}|^{2}}{2 \pi} \frac{\hbar K_{n}}{M}. \label{Jntr}
\end{align}
Thus the corresponding coefficients of reflection and transmission are
\begin{align}
  j_{n}^{re} &= \frac{|J_{n}^{re}|}{J_{0}^{in}} = |\alpha_{n}|^{2} \frac{K_{n}}{K_{0}}, \label{jnre}\\
  j_{n}^{tr} &= \frac{J_{n}^{tr}}{J_{0}^{in}} = |\beta_{n}|^{2} \frac{K_{n}}{K_{0}}. \label{jntr}
\end{align}
Note that the conservation of probability implies
\begin{equation}
  j_{t} = \sum_{n} \left(j_{n}^{re} + j_{n}^{tr}\right) = 1.
  \label{conser}
\end{equation}

The boundary conditions at $x_{1}=0$ and $x_{2}=0$ require
\begin{align}
  \begin{cases}
    \hspace{3.4cm} \Psi_{K_{0},l}^{+}|_{x_{1} \rightarrow 0^{+}} &=
    \Psi_{K_{0},l}^{+}|_{x_{1} \rightarrow 0^{-}} ,\\
    \pdv{x_{1}} {\Psi_{K_{0},l}^{+}|_{x_{1}
        \rightarrow 0^{+}}} - \pdv{x_{1}}
    {\Psi_{K_{0},l}^{+}|_{x_{1} \rightarrow 0^{-}}} &= \frac{2m_{1}
    \gamma_{1}}{\hbar^{2}} \Psi_{K_{0},l}^{+}|_{x_{1} \rightarrow 0^{+}},\\
    \hspace{3.4cm} \Psi_{K_{0},l}^{+}|_{x_{2} \rightarrow 0^{+}} &=
    \Psi_{K_{0},l}^{+}|_{x_{2} \rightarrow 0^{-}} ,\\
    \pdv{x_{2}} {\Psi_{K_{0},l}^{+}|_{x_{2}
        \rightarrow 0^{+}}} - \pdv{x_{2}}
    {\Psi_{K_{0},l}^{+}|_{x_{2} \rightarrow 0^{-}}} &= \frac{2m_{2}
    \gamma_{2}}{\hbar^{2}} \Psi_{K_{0},l}^{+}|_{x_{2} \rightarrow 0^{+}},
  \end{cases}\label{condition}
\end{align}
Concretely speaking, the continuity of the wave function at $L_{1}$
(see Fig.~\ref{schematic}) gives
\begin{align}
  & e^{iK_{0} r_{1} x_{1}}
    \psi_{l} (-x_{1}) + \sum_{n} \alpha_{n} e^{-iK_{n}r_{1} x_{1}}
    \psi_{n} (-x_{1}) \nonumber\\
  & = \sum_{n} \mu_{n} e^{iK_{n} r_{1} x_{1}} \psi_{n} (-x_{1}) +
    \sum_{n} \nu_{n} e^{-iK_{n} r_{1} x_{1}} \psi_{n}
    (-x_{1}),\nonumber\\
  & \hspace{6cm} (x_{1}<0)
    \label{l1continue}
\end{align}
and the discontinuity of the derivative of wave function at $L_{1}$ gives
\begin{align}
  & \frac {2m_{2} \gamma_{2}}{\hbar^{2}} \left[ \sum_{n} \mu_{n}
    e^{iK_{n} r_{1} x_{1}} \psi_{n} (-x_{1}) + \sum_{n} \nu_{n}
    e^{-iK_{n} r_{1} x_{1}} \psi_{n} (-x_{1}) \right] \nonumber\\
  & = \sum_{n} \mu_{n} \left[ e^{iK_{n} r_{1} x_{1}} \psi_{n}^{'}
    (-x_{1}) + i K_{n} r_{2}  e^{iK_{n} r_{1} x_{1}} \psi_{n} (-x_{1})
    \right]\nonumber\\
  & \quad{} + \sum_{n} \nu_{n} \left[e^{-iK_{n} r_{1} x_{1}}
    \psi_{n}^{'} (-x_{1}) - iK_{n} r_{2} e^{-iK_{n} r_{1} x_{1}}
    \psi_{n} (-x_{1}) \right]\nonumber\\
  & \quad{} - \left[e^{iK_{0} r_{1} x_{1}} \psi_{0}^{'} (-x_{1}) +
    e^{iK_{0} r_{1} x_{1}}iK_{0} r_{2} \psi_{0} (-x_{1}) \right]
    \nonumber\\
  & \quad{} - \sum_{n} \alpha_{n} \left[e^{-iK_{n} r_{1} x_{1}}
    \psi_{n}^{'} (-x_{1}) - iK_{n}r_{2} e^{-iK_{n} r_{1} x_{1}}
    \psi_{n} (-x_{1}) \right]\nonumber\\
  & \hspace{6cm} (x_{1}<0).\label{l1discontinue}
\end{align}
Multiplying Eq.~\ref{l1continue} and Eq.~\ref{l1discontinue} with
$\psi_{m}$ and integrating from $0$ to $\infty$ leads to
\begin{eqnarray}
  && c_{ml}^{2} (-iK_{0} r_{1}) + \sum_{n} \alpha_{n} c_{mn}^{2}
     (iK_{n} r_{1})\nonumber\\
  && = \sum_{n} \mu_{n} c_{mn}^{2} (-iK_{n} r_{1}) + \sum_{n} \nu_{n}
     c_{mn}^{2}(iK_{n} r_{1})\label{condition1}
\end{eqnarray}
and
\begin{align}
  & \frac{2m_{2} \gamma_{2}}{\hbar^{2}} \left[ \sum_{n} \mu_{n}
    c_{mn}^{2} (-iK_{n} r_{1}) + \sum_{n} \nu_{n} c_{mn}^{2}(iK_{n}
    r_{1})\right] \nonumber\\
  & = \sum_{n} \mu_{n} \left[ iK_{n} r_{2} c_{mn}^{2} (-iK_{n}r_{1}) +
    d_{mn}^{2} (-iK_{n} r_{1}) \right]\nonumber\\
  & \quad{} + \sum_{n} \nu_{n} \left[ -iK_{n} r_{2} c_{mn}^{2} (iK_{n}
    r_{1}) + d_{mn}^{2} (iK_{n} r_{1}) \right]\nonumber\\
  & \quad{} - \left[ iK_{0} r_{2} c_{ml}^{2} (-iK_{0} r_{1}) +
    d_{ml}^{2} (-iK_{0} r_{1}) \right]\nonumber\\
  & \quad{} - \sum_{n} \alpha_{n} \left[ -iK_{n} r_{2} c_{mn}^{2}
    (iK_{n} r_{1}) + d_{mn}^{2} (iK_{n} r_{1}) \right].
    \label{condition2}
\end{align}
Similarly, the boundary conditions at $L_{2}$, $L_{3}$ and $L_{4}$
take the form
\begin{align}
  & \sum_{n} \mu_{n} c_{mn}^{2} (iK_{n} r_{2}) + \sum_{n} \nu_{n}
    c_{mn}^{2} (-iK_{n} r_{2})\nonumber\\
  & = \sum_{n} \beta_{n} c_{mn}^{2} (iK_{n} r_{2}),\label{condition3}
\end{align}
\begin{align}
  &\frac{2m_{1} \gamma_{1}}{\hbar^{2}} \sum_{n} \beta_{n} c_{mn}^{2}
    (ik_{n} r_{2}) \nonumber\\
  & = \sum_{n} \beta_{n} \left[ iK_{n} r_{1} c_{mn}^{2} (iK_{n} r_{2})
    - d_{mn}^{2} (iK_{n} r_{2}) \right] \nonumber\\
  & \quad{} -\sum_{n} \mu_{n} \left[ iK_{n} r_{1} c_{mn}^{2} (iK_{n}
    r_{2}) - d_{mn}^{2} (iK_{n} r_{2}) \right] \nonumber\\
  & \quad{} - \sum_{n} \nu_{n} \left[ -iK_{n}r_{1} c_{mn}^{2} (-iK_{n}
    r_{2}) - d_{mn}^{2} (-iK_{n} r_{2}) \right] ,\label{condition4}
\end{align}
\begin{align}
  & c_{ml}^{1} (iK_{0} r_{2}) + \sum_{n} \alpha_{n} c_{mn}^{1}
    (-iK_{n} r_{2})\nonumber\\
  & = \sum_{n} \xi_{n} c_{mn}^{1} (iK_{n} r_{2}) + \sum_{n} \eta_{n}
    c_{mn}^{1} (-iK_{n} r_{2}),\label{condition5}
\end{align}
\begin{align}
  & \frac{2m_{1} \gamma_{1}}{\hbar^{2}} \left[ \sum_{n} \xi_{n}
    c_{mn}^{1}(iK_{n} r_{2}) + \sum_{n} \eta_{n} c_{mn}^{1}(-ik_{n}
    r_{2}) \right]\nonumber\\
  & = \sum_{n} \xi_{n} \left[ iK_{n} r_{1} c_{mn}^{1}(iK_{n} r_{2}) -
    d_{mn}^{1}(iK_{n} r_{2}) \right]\nonumber\\
  & \quad{} + \sum_{n} \eta_{n} \left[ -iK_{n} r_{1} c_{mn}^{1}
    (-iK_{n} r_{2}) - d_{mn}^{1} (-iK_{n} r_{2}) \right]\nonumber\\
  & \quad{} - \left[ iK_{0} r_{1} c_{ml}^{1} (iK_{0} r_{2}) -
    d_{ml}^{1} (iK_{0} r_{2}) \right]\nonumber\\
  & \quad{} - \sum_{n} \alpha_{n} \left[ -iK_{n} r_{1} c_{mn}^{1}
    (-iK_{n} r_{2}) - d_{mn}^{1} (-iK_{n} r_{2})
    \right],\label{condition6}
\end{align}
\begin{align}
  & \sum_{n} \xi_{n} c_{mn}^{1} (-iK_{n} r_{1}) + \sum_{n} \eta_{n}
    c_{mn}^{1} (iK_{n} r_{1})\nonumber\\
  & = \sum_{n} \beta_{n} c_{mn}^{1} (-iK_{n} r_{1}),\label{condition7}
\end{align}
and
\begin{align}
  & \frac{2m_{2} \gamma_{2}}{\hbar^{2}} \sum_{n} \beta_{n} c_{mn}^{1}
    (-iK_{n} r_{1})\nonumber\\
  & = \sum_{n} \beta_{n} \left[ iK_{n} r_{2} c_{mn}^{1} (-iK_{n}
    r_{1}) + d_{mn}^{1} (-iK_{n} r_{1}) \right]\nonumber\\
  & \quad{} - \sum_{n} \xi_{n} \left[ iK_{n} r_{2} c_{mn}^{1} (-iK_{n}
    r_{1}) + d_{mn}^{1} (-iK_{n} r_{1}) \right]\nonumber\\
  & \quad{} - \sum_{n} \eta_{n} \left[ -iK_{n} r_{2} c_{mn}^{1}
    (iK_{n} r_{1}) + d_{mn}^{1} (iK_{n} r_{1})
    \right], \label{condition8}
\end{align}
where
$c_{mn}^{1}(q) = \int_{-\infty}^{0} \psi_{m}(x) e^{qx} \psi_{n}(x)
\mathbf{d}x$, $c_{mn}^{2}(q) = \int_{0}^{\infty} \psi_{m}(x) e^{qx}
\psi_{n}(x) \mathbf{d}x$,
$d_{mn}^{1}(q) = \int_{-\infty}^{0} \psi_{m}(x) e^{qx} \psi_{n}^{'}(x)
\mathbf{d}x$ and
$d_{mn}^{2}(q) = \int_{0}^{\infty} \psi_{m}(x) e^{qx} \psi_{n}^{'}(x)
\mathbf{d}x$.

Eqs.\ref{condition1}$-$\ref{condition8} determine $\Psi_{K_{0},l}^{+}$. In
the following we will analyze the scattering behavior of this
two-particle composite system.

\section{Results and Analysis}

In this section, we calculate how the population of excited states
changes with different parameters, and analyze the physical picture
behind it. In the following calculation, we  set $\hbar=1$, $M=2$
and select the incident internal state  to be the ground state or
the first excited state, i.e., $l=0$ or $1$.

\begin{figure}[htbp]
  \centering %
  \subfloat[][]{\label{nab} \includegraphics[width=7.5cm]{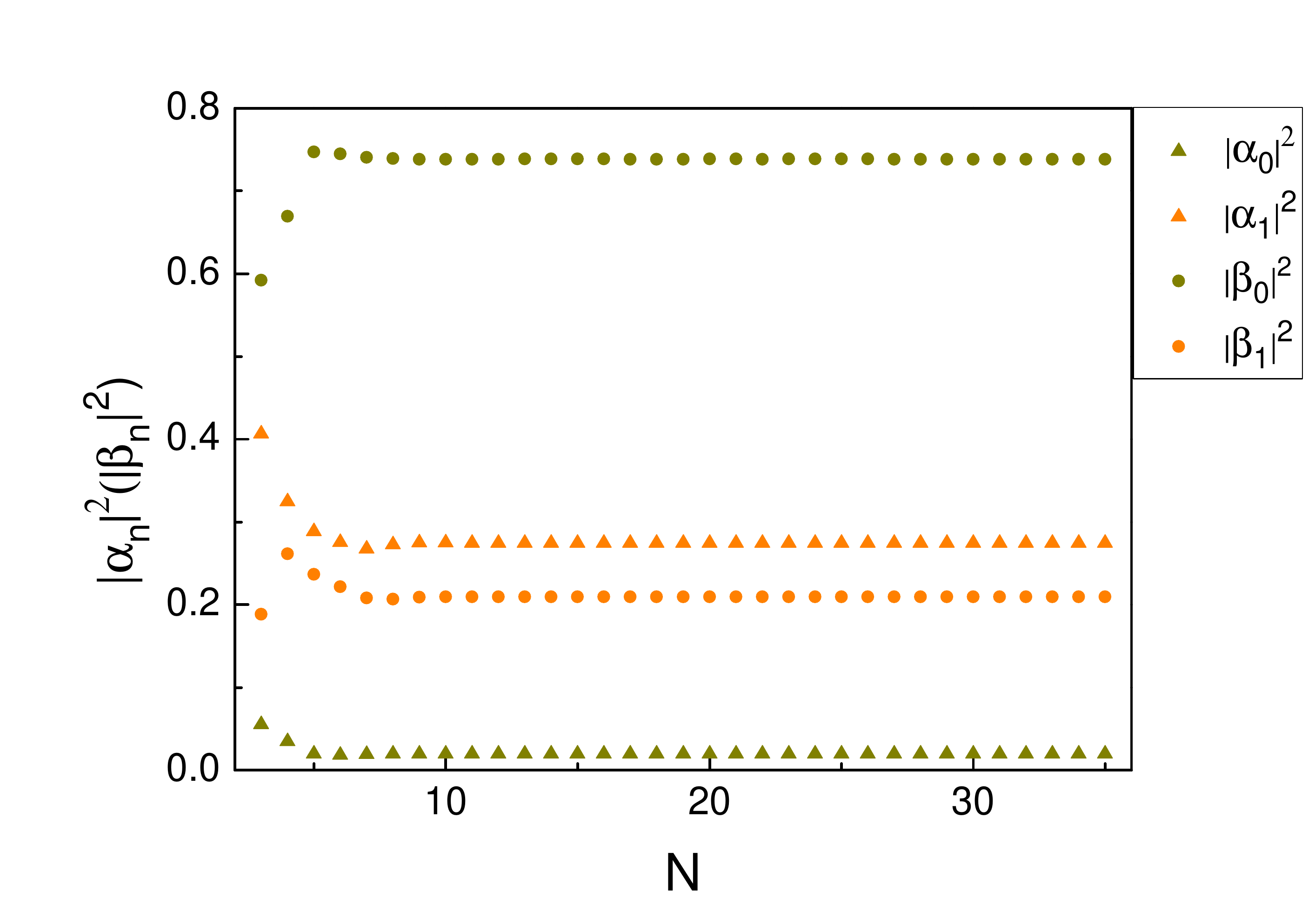}}
  \quad
  \subfloat[][]{\label{nj} \includegraphics[width=7.5cm]{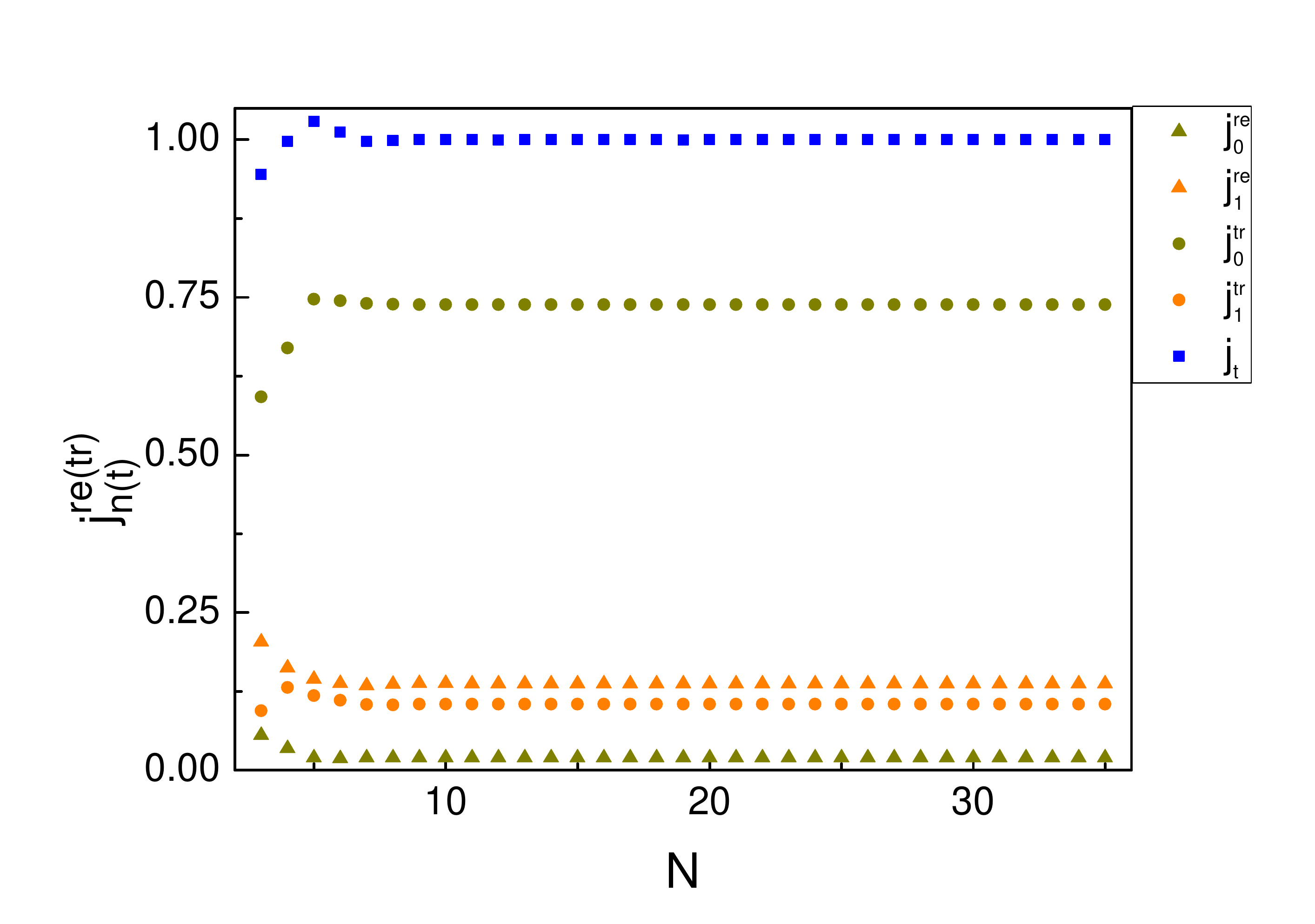}}
  \caption{(Color online) The amplitudes $\alpha_{n},\beta_{n}$ (a)
    and reflection/transmission coefficients $j_{n}$ (b) via the total modes $N$ with
    parameters $K_{0}=4$, $\omega=3$, $\gamma_{1}=1$, $\gamma_{2}=0$
    and $n_{c}=1$.}~\label{nabj}
\end{figure}

\begin{figure}[htbp]
  \centering %
  \subfloat[][]{\label{gaja} \includegraphics[width=7.5cm]{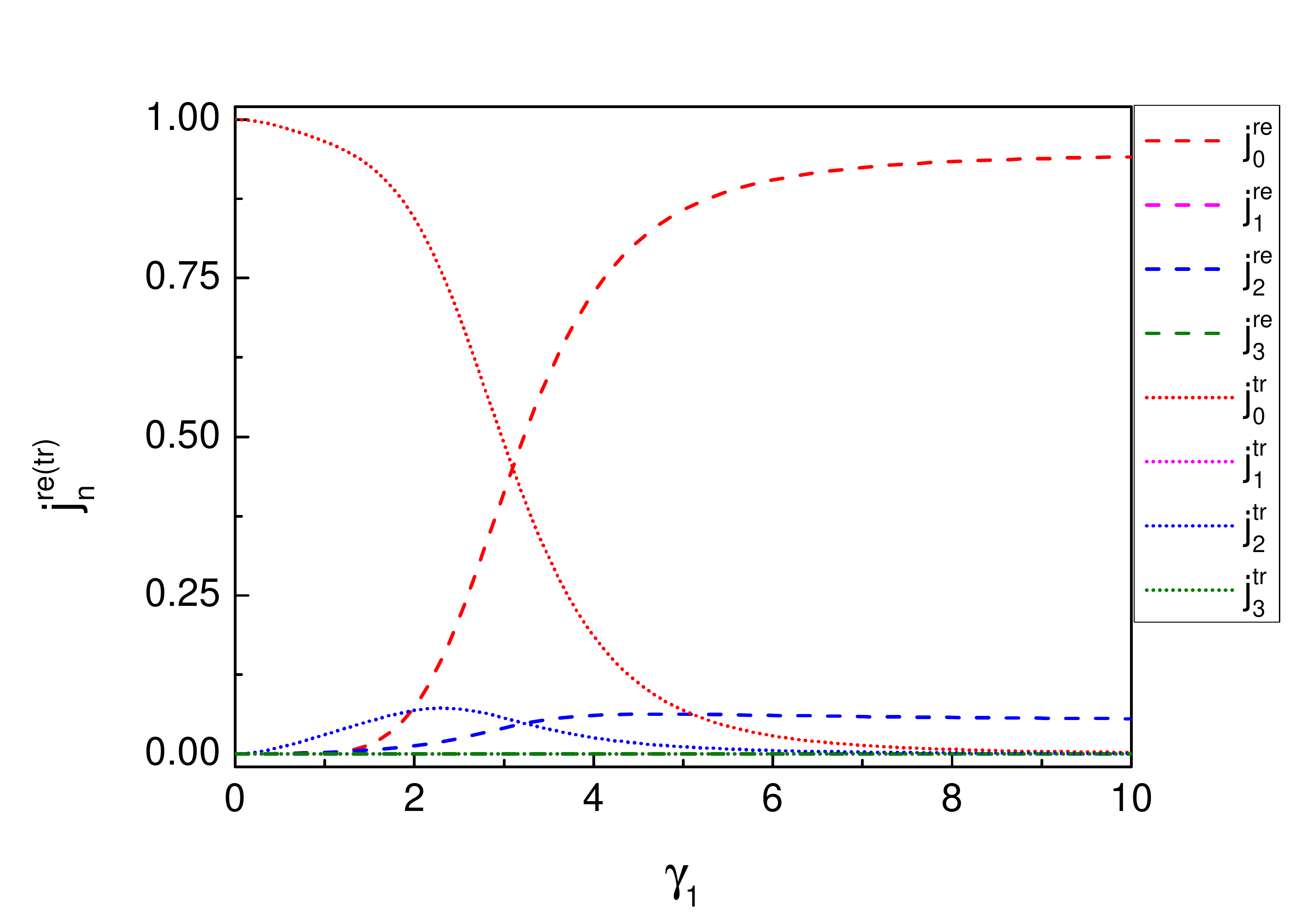}}
  \quad
  \subfloat[][]{\label{gajb} \includegraphics[width=7.5cm]{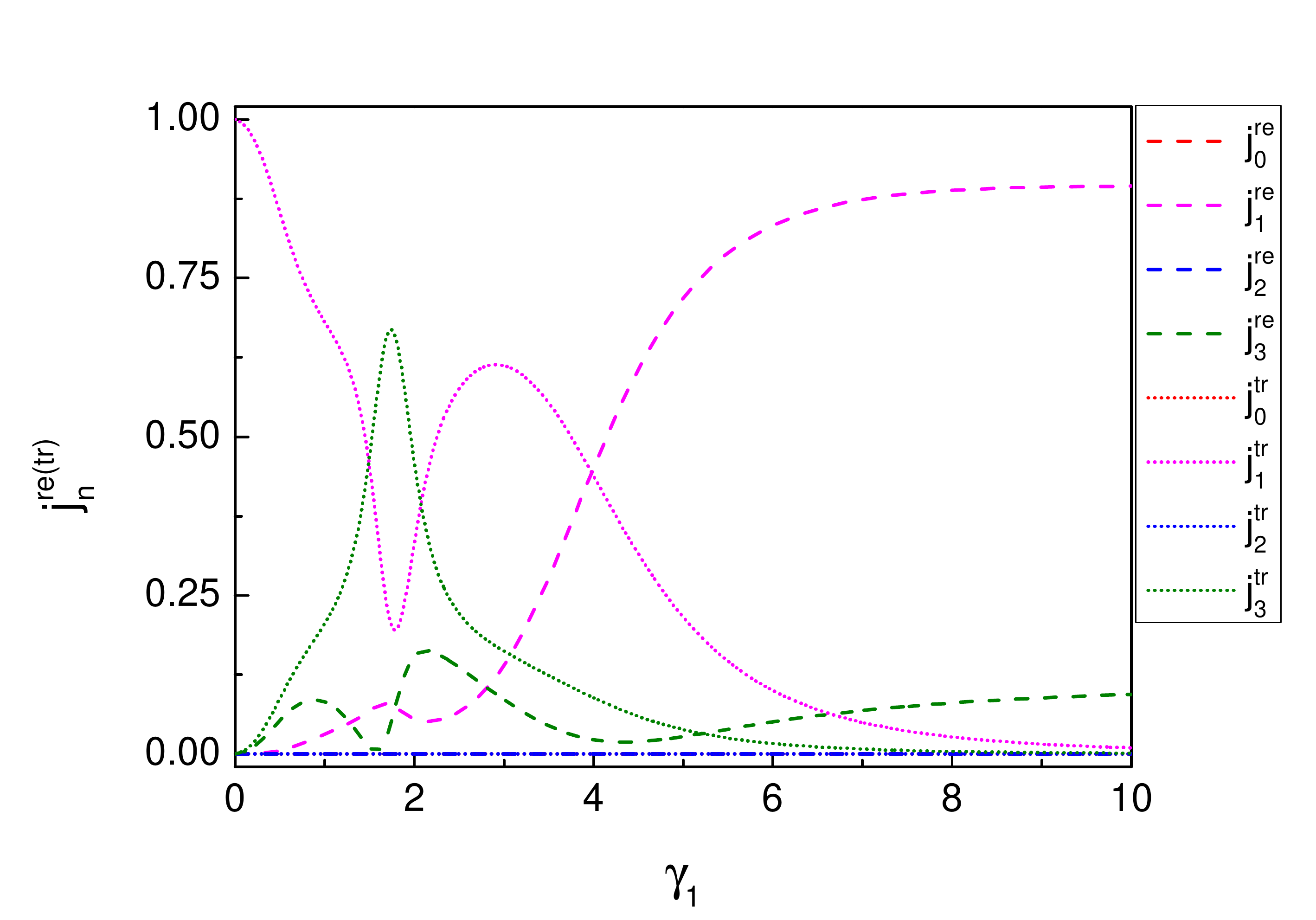}}\quad
  \subfloat[][]{\label{gajc} \includegraphics[width=7.5cm]{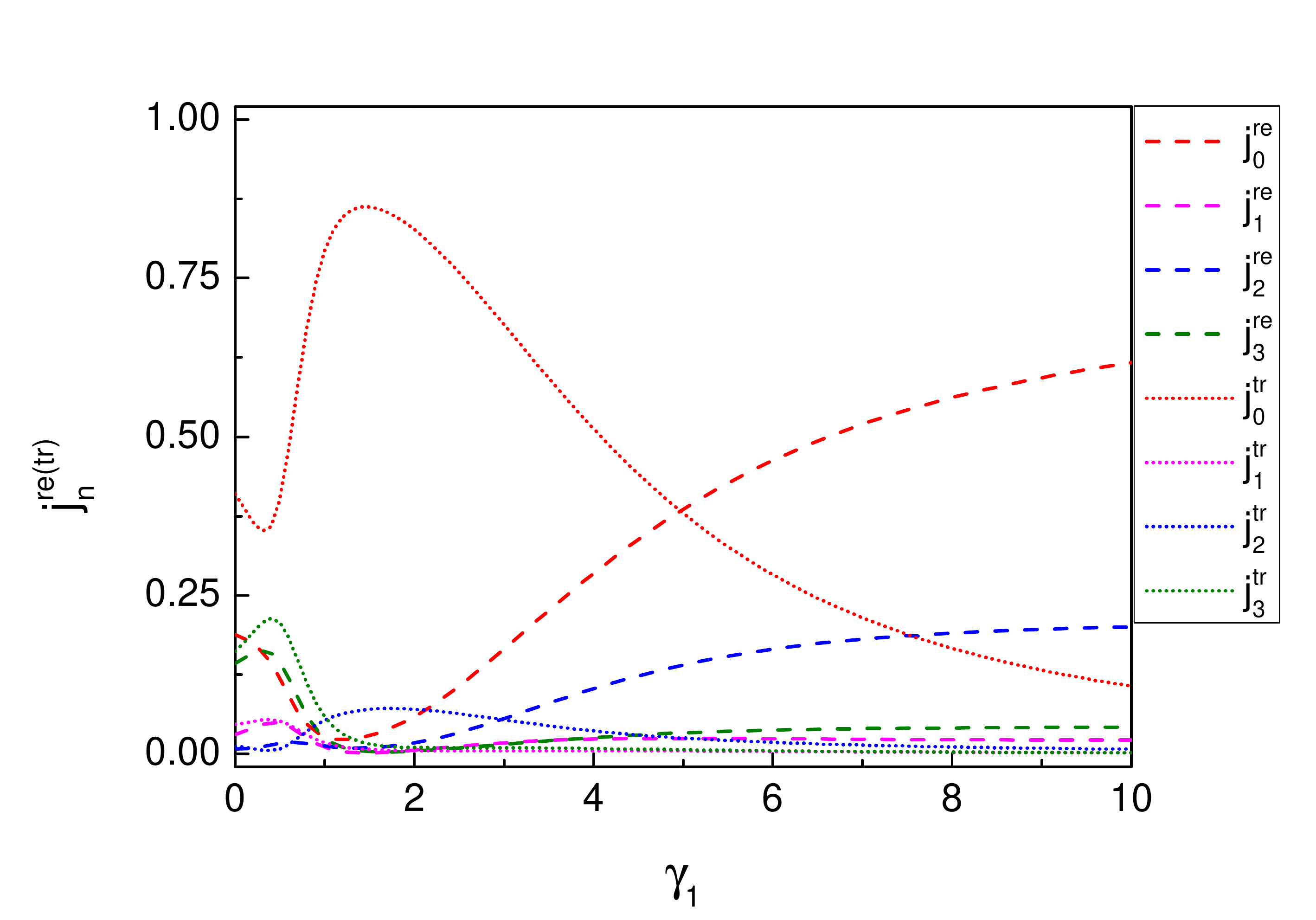}}
  \caption{(Color online) The reflection/transmission coefficients $j_{n}$ via the
    potential strength $\gamma_{1}$ with parameters (a) $m_{1}=1$, $K_{0}=5.2$,
    $\omega=2$, $\gamma_{2}=\gamma_{1},l=0$; (b) $m_{1}=1$, $K_{0}=4.5$,
    $\omega=2$, $\gamma_{2}=\gamma_{1},l=1$;  (c) $m_{1}=1.1$, $K_{0}=5.2$,
    $\omega=2$, $\gamma_{2}=2,l=0$.}~\label{gaj}
\end{figure}

From Eq.~\ref{energyconser}, it's reasonable to believe that: (1) the
$n$-th excited internal state may become populated if its energy is less
than the energy of the incident wave
$(n + 1/2)\hbar \omega \leq \hbar^{2} K_{0}^{2}/2 M + (l + 1/2)\hbar \omega$; (2) the
highest internal energy level excited by the potential should be
$n_{c}=\lfloor\hbar K_{0}^{2}/(2M\omega)\rfloor+l$, where $\lfloor a
\rfloor$ is the maximum integer less than or equal to $a$. However it is
worthy to point out that in Eqs.\ref{condition1}$-$\ref{condition8}, the levels above $n_{c}$ should also be taken into account
to ensure the conservation of probability is satisfied, thus $K_{n}=i\sqrt{2nM\omega-K_{0}^{2}}$ for $n>n_{c}$. These
correspond to the states whose internal energies are sufficiently high
while the center-of-mass energies being negative. From the point of
physics, these sates only exist in the scattering region since they
decay rapidly as $|X|\rightarrow \infty$. In fact, it's enough to take
only several internal modes above $n_{c}$ into account.
Fig.~\ref{nabj} shows that the results are stable with the increasing
of the mode number, indicating that the modes well above $n_{c}$ have
little effect on the population of the actual states.

\begin{figure}[htbp]
  \centering %
  \subfloat[][]{\label{kja} \includegraphics[width=7.5cm]{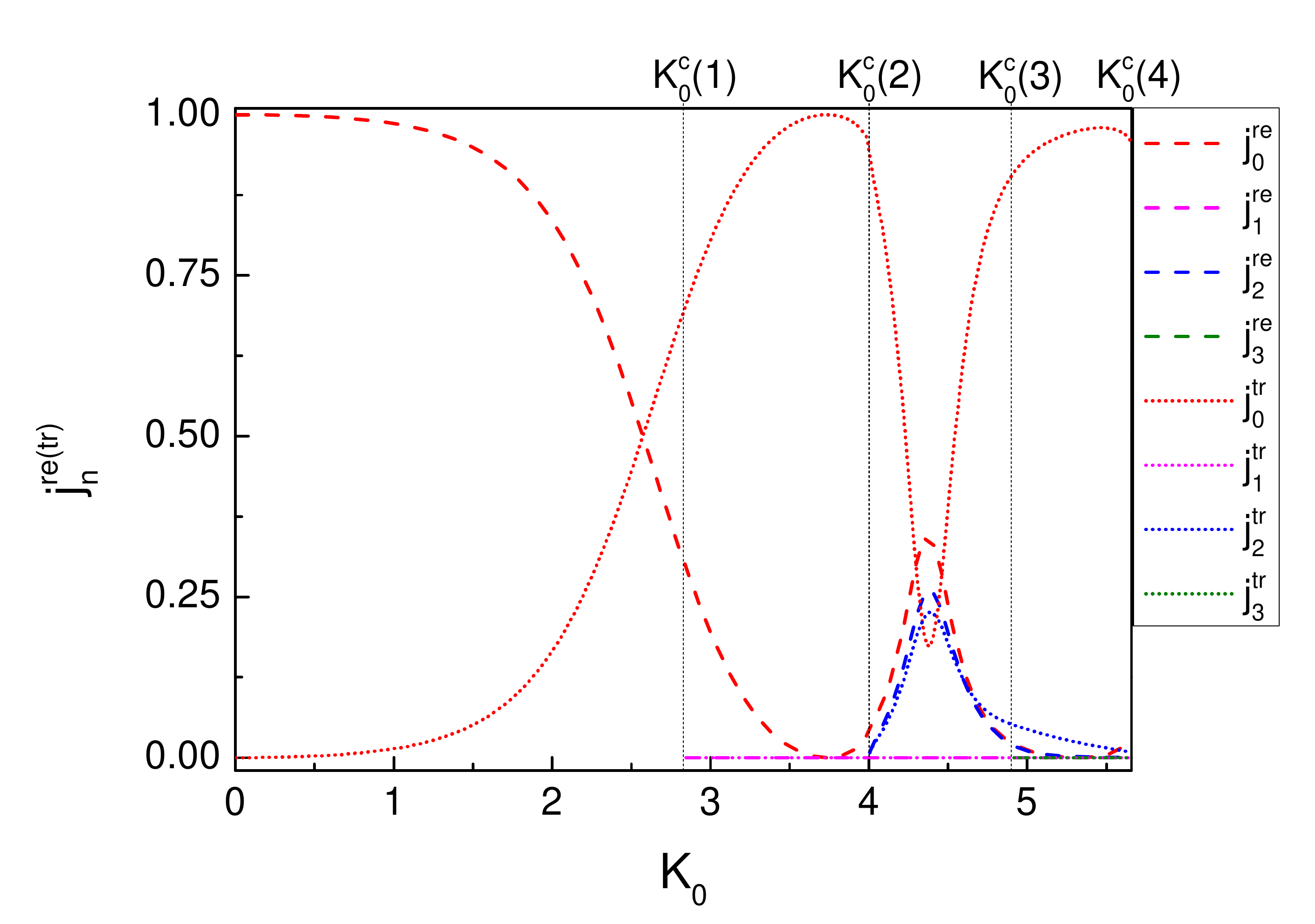}}
  \quad
  \subfloat[][]{\label{kjb} \includegraphics[width=7.5cm]{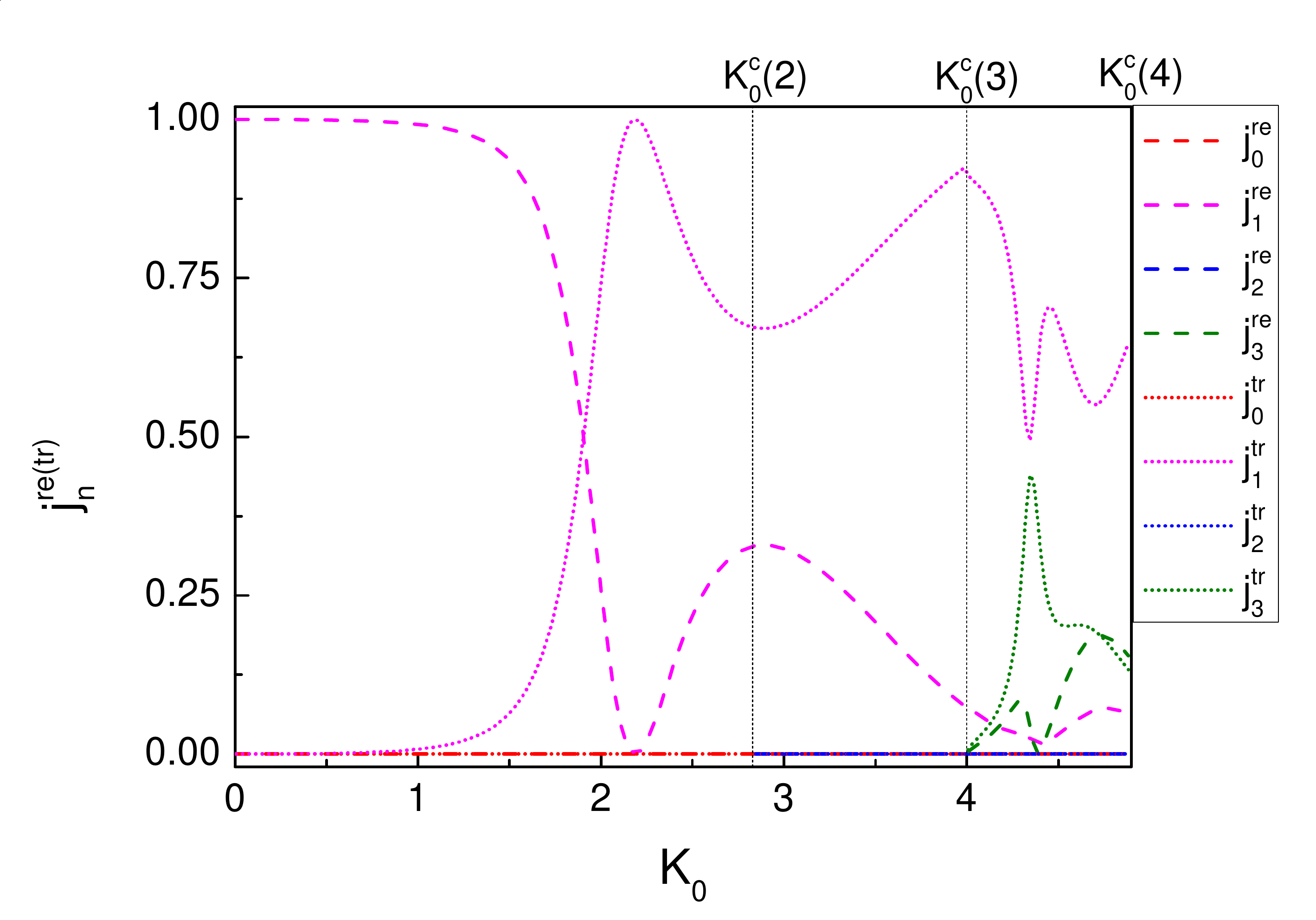}}
  \quad
  \subfloat[][]{\label{kjc} \includegraphics[width=7.5cm]{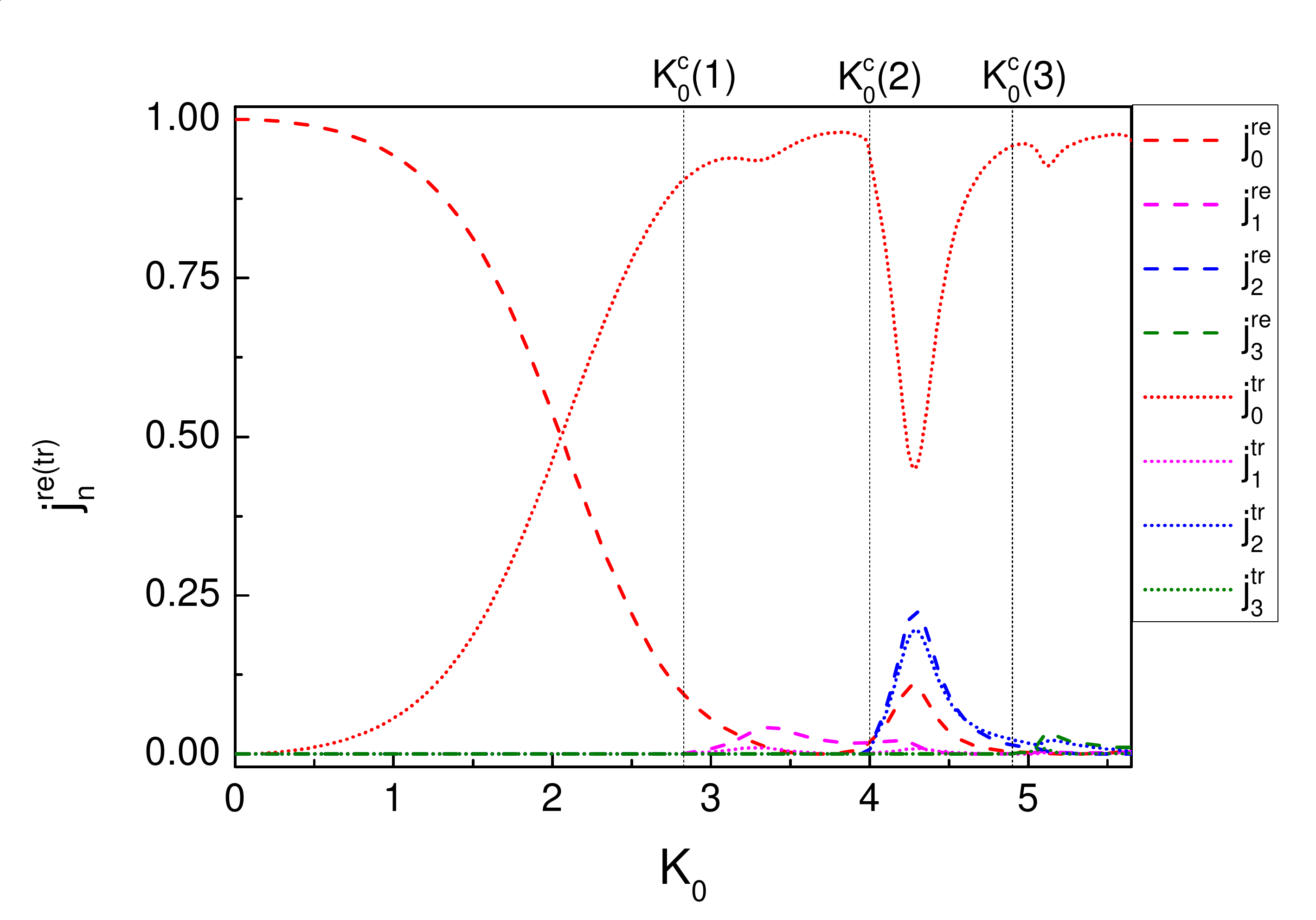}}
  \caption{(Color online) The reflection/transmission coefficients
    $j_{n}$ via the incident momentum of the center-of-mass $K_{0}$
    with parameters (a) $m_{1}=1$, $\omega=2$,
    $\gamma_{1}=\gamma_{2}=1,l=0$; (b) $m_{1}=1$, $\omega=2$,
    $\gamma_{1}=\gamma_{2}=1,l=1$; (c) $m_{1}=1.1$, $\omega=2$,
    $\gamma_{1}=0.8,\gamma_{2}=0.5,l=0$.}~\label{kj}
\end{figure}

The influences of barrier heights $\gamma_{1,2}$ are different for the
reflected and transmitted part of the wave. As a consequence, the
internal modes become populated differently. As shown in
Fig.~\ref{gaja} and Fig.~\ref{gajb}, in the limit
$\gamma_{1}=\gamma_{2}=0$, the incident wave $\phi_{K_{0},l}$
transmits directly without reflection and excitation ($j_{l}^{tr}=1$).
For $l=0$ (Fig.~\ref{gaja}), with the increasing of $\gamma_{1}$, the
transmission coefficient for ground mode $j_{0}^{tr}$ decreases
monotonously while the reflection coefficient $j_{0}^{re}$ increases. It should be stressed
that, at the quite beginning of $\gamma_{1}$, the
reflection/transmission coefficients for excited states $j_{2}^{re(tr)}$
increase, indicating that a certain height of potential is needed to
excite the upper states. With the further increasing of $\gamma_{1}$,
$j_{n}^{re}$ increase and $j_{n}^{tr}$ decrease, only reflected
components remains for the potential high enough.

We want to emphasize that the influence of a symmetric Hamiltonian( i.e.,
$m_{1}=m_{2},\gamma_{1}=\gamma_{2}$) is
greater on the odd-parity internal states than on the even-parity ones.
The curves of $j_{1,3}$ in Fig.~\ref{gajb} have obviously characteristics
"peak" and "valley" which are not available for $j_{0,2}$ in Fig.~\ref{gaja}.
Moreover, Fig.~\ref{gaj} illustrates the dependence of the internal
excitation on the symmetry of Hamiltonian. The internal degree of
freedom can only be excited to the states whose parity is same as that
of the incident state for a symmetric Hamiltonian, (Fig.~\ref{gaja},
Fig.~\ref{gajb}), contrasting to that all the states $n\leq n_{c}$
could be excited for a asymmetric Hamiltonian $m_{1}\neq m_{2}$,$\gamma_{1}\neq \gamma_{2}$ (Fig.~\ref{gajc}).

From Eq.~\ref{energyconser}, the critical incident momentums which
could excite $n$-th internal states read
\begin{equation}
  K_{0}^{c}(n) = \sqrt{2(n-l) M \omega / \hbar},
  \label{k0c}
\end{equation}
if $K_{0}>K_{0}^{c}(n)$, the $n$-th internal state may be excited.
This is the energy condition to excite internal states. Fig.~\ref{kj}
displays how reflection and transmission coefficients $j_{n}$ change with $K_{0}$, the
vertical lines label $K_{0}^{c}$ to excite $n=1,2,3$ states. For $l=0$
(Fig.~\ref{kja}), when $K_{0}$ is not too large, the internal degree of
freedom is in the ground state, the reflection (transmission) coefficient
decreases (increases) with the increasing of $K_{0}$, but for
$K_{0}=K_{0}^{c}(2)$, the $n=2$ internal state becomes populated,
leading to the non-monotonic change of $j_{0}$ with $K_{0}$. However,
when $K_{0}>K_{0}^{c}(1)$ and $K_{0}>K_{0}^{c}(3)$, $n=1$ and $n=3$
states are non-excited because of the symmetry although energy
condition Eq.~\ref{k0c} is satisfied. Similar behaviors were also
found for $l=1$. Additionally, the reflection and transmission coefficients
of the excited internal state for $l=1$( $j_{3}^{re(tr)}$ in Fig.~\ref{kjb}) have obviously
characteristics "peak" and "valley" whereas those for $l=0$ ( $j_{2}^{re(tr)}$ in Fig.~\ref{kja})
have only a "peak", confirming the greater influence of a symmetric Hamiltonian on the odd-parity state.
In Fig.~\ref{kjc}, since $\gamma_{1}\neq\gamma_{2}$ and $m_{1}\neq m_{2}$
the $n$-th state can be excited as long as $K_{0}>K_{0}^{c}(n)$.
Furthermore, compared with the other excited states, the modes $n=2$
is mainly populated because the symmetry is just damaged slightly for this set of parameters.

It is natural that the internal excitation will affect the reflected and transmitted components,
however what we emphasize is that the internal degree of freedom will significantly changes the
the reflection and transmission coefficients even without internal excitation. See Fig.~\ref{kjb},
when $K_{0} < K_{0}^{c}(3)$, although no excited states are populated besides the
incident $l=1$ state, the curves of $j_{1}^{re(tr)}$ present non-monotonic dependence on $K_{0}$.

\begin{figure}[htbp]
  \centering %
  \subfloat[][]{\label{omja} \includegraphics[width=7.5cm]{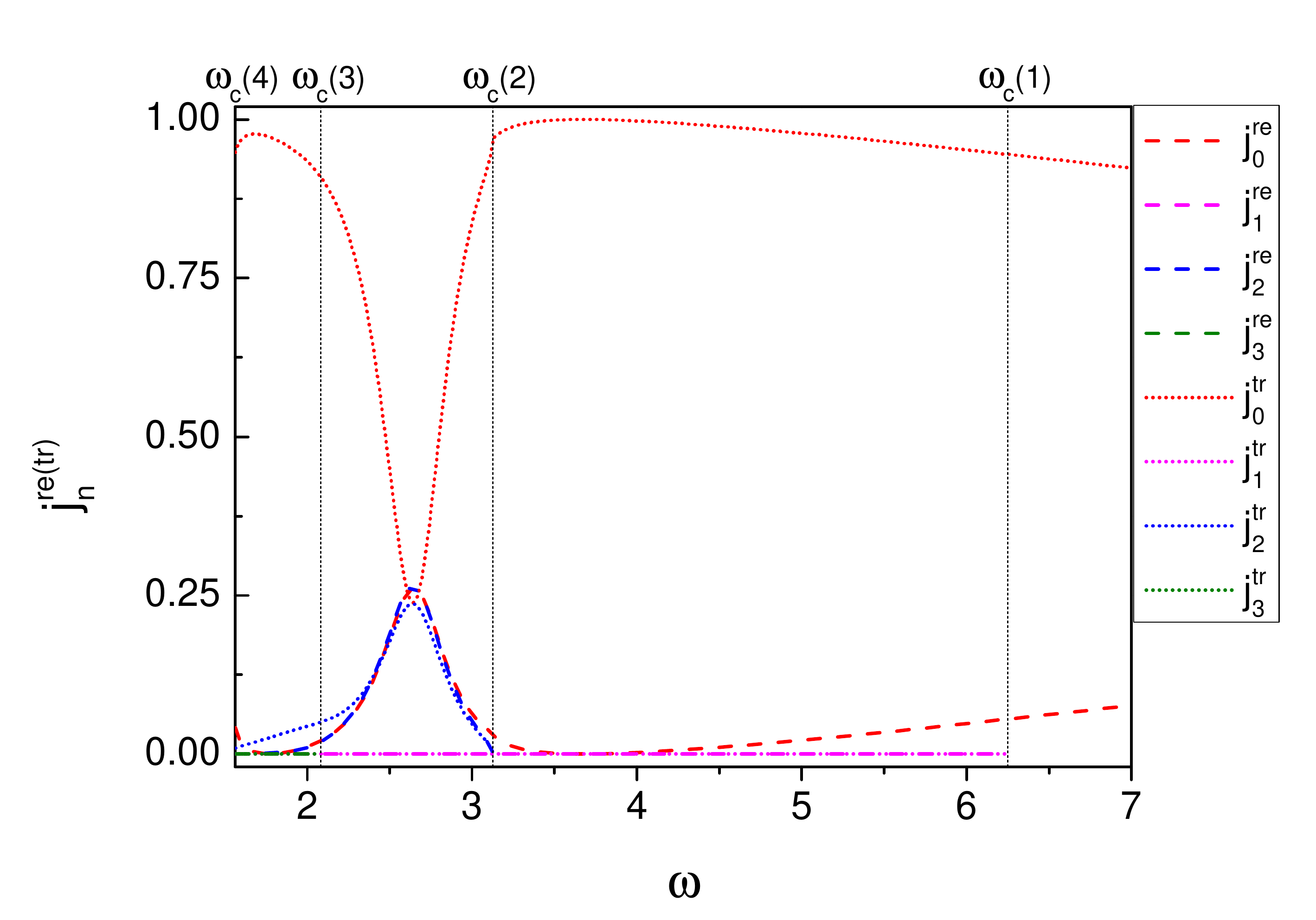}}
  \quad
  \subfloat[][]{\label{omjb} \includegraphics[width=7.5cm]{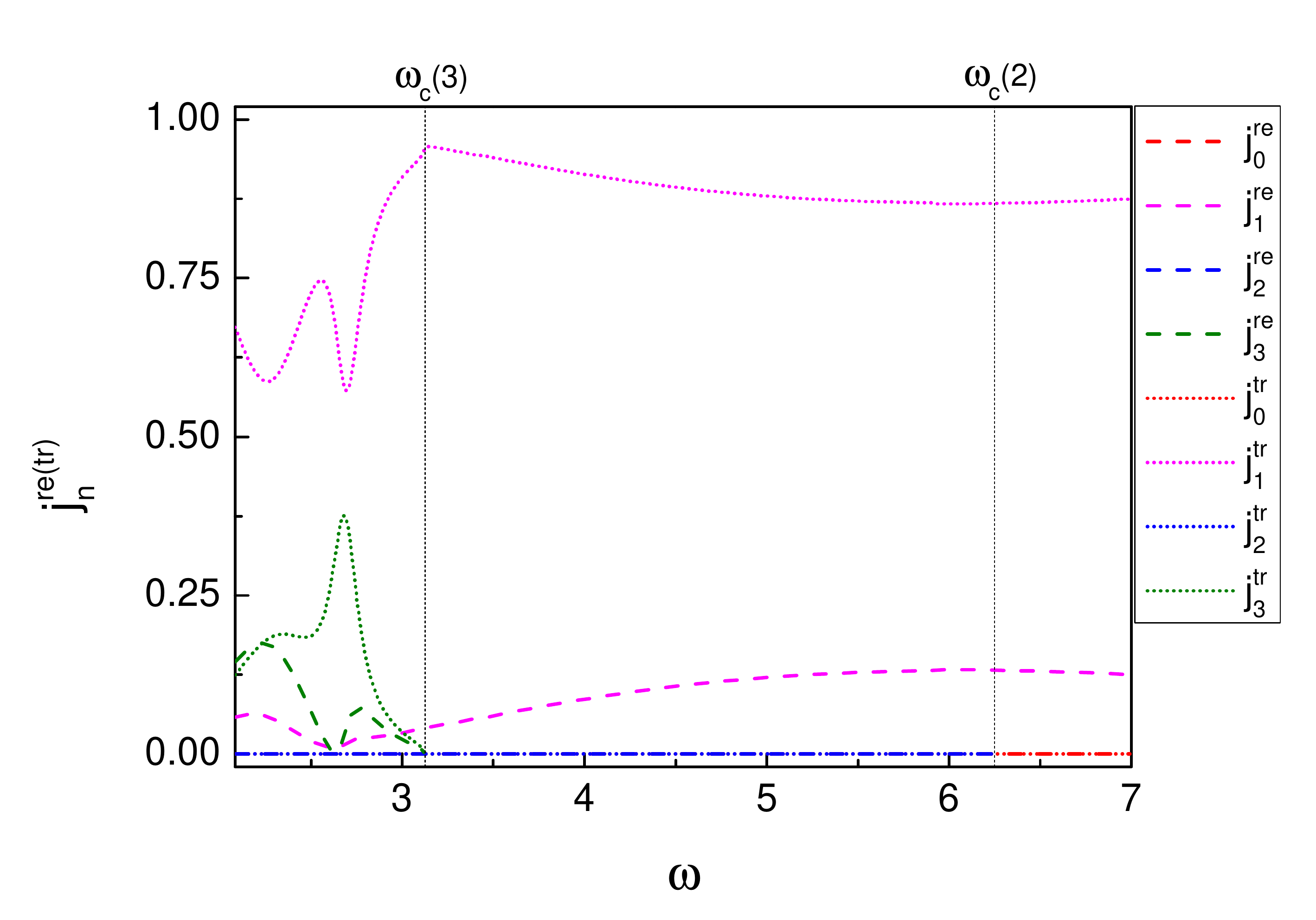}}\quad
  \subfloat[][]{\label{omjc} \includegraphics[width=7.5cm]{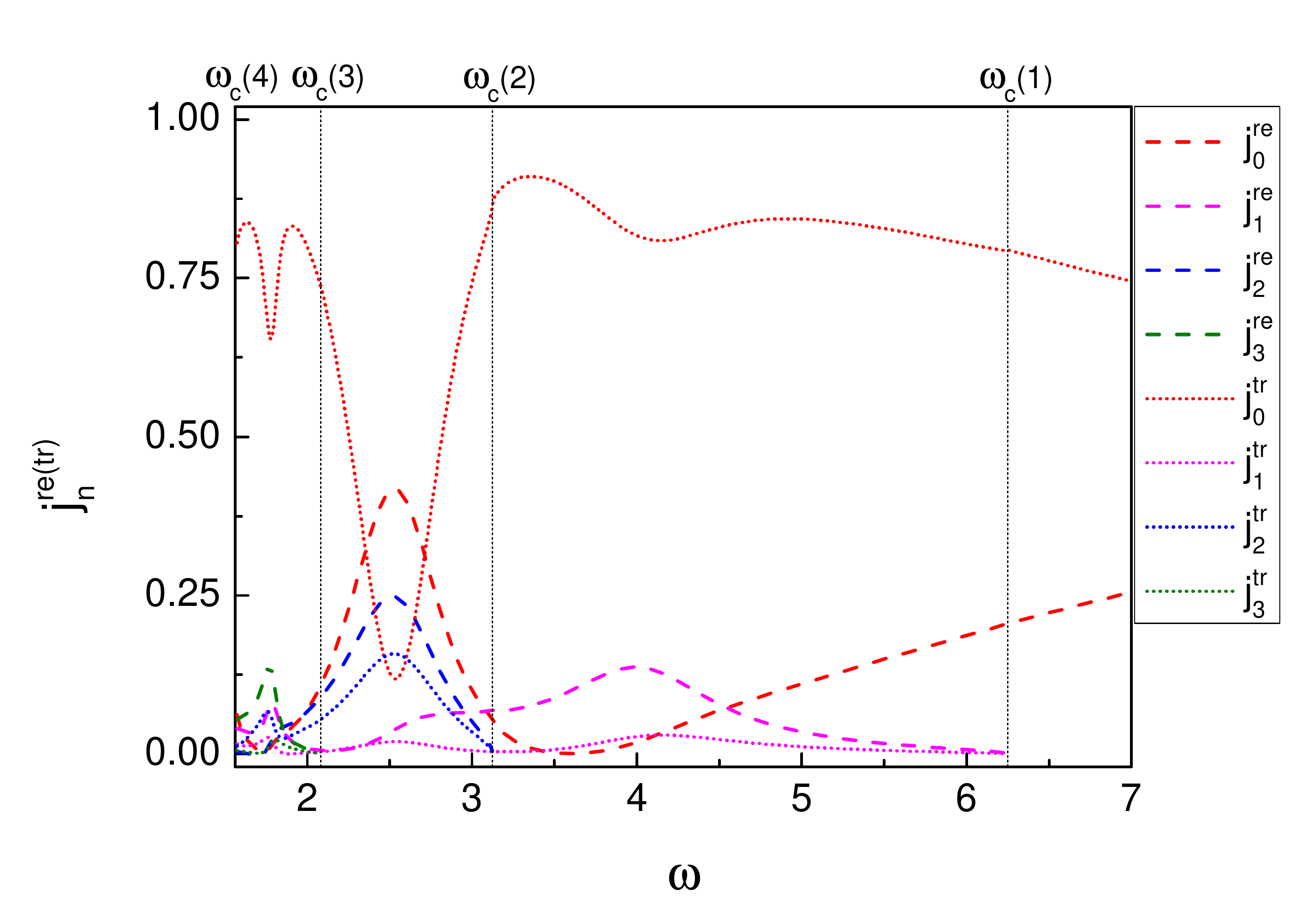}}
  \caption{(Color online) The  reflection/transmission coefficients $j_{n}$ via the
    potential strength $\omega$ with parameters (a) $m_{1}=1$,
    $K_{0}=5$, $\gamma_{1}=\gamma_{2}=1,l=0$; (b) $m_{1}=1$,
    $K_{0}=5$, $\gamma_{1}=\gamma_{2}=1,l=1$; (c) $m_{1}=1.1$,
    $K_{0}=5$, $\gamma_{1}=2,\gamma_{2}=1,l=0$.}~\label{omj}
\end{figure}

\begin{figure}[htbp]
  \centering %
  \subfloat[][]{\label{omj1} \includegraphics[width=7.5cm]{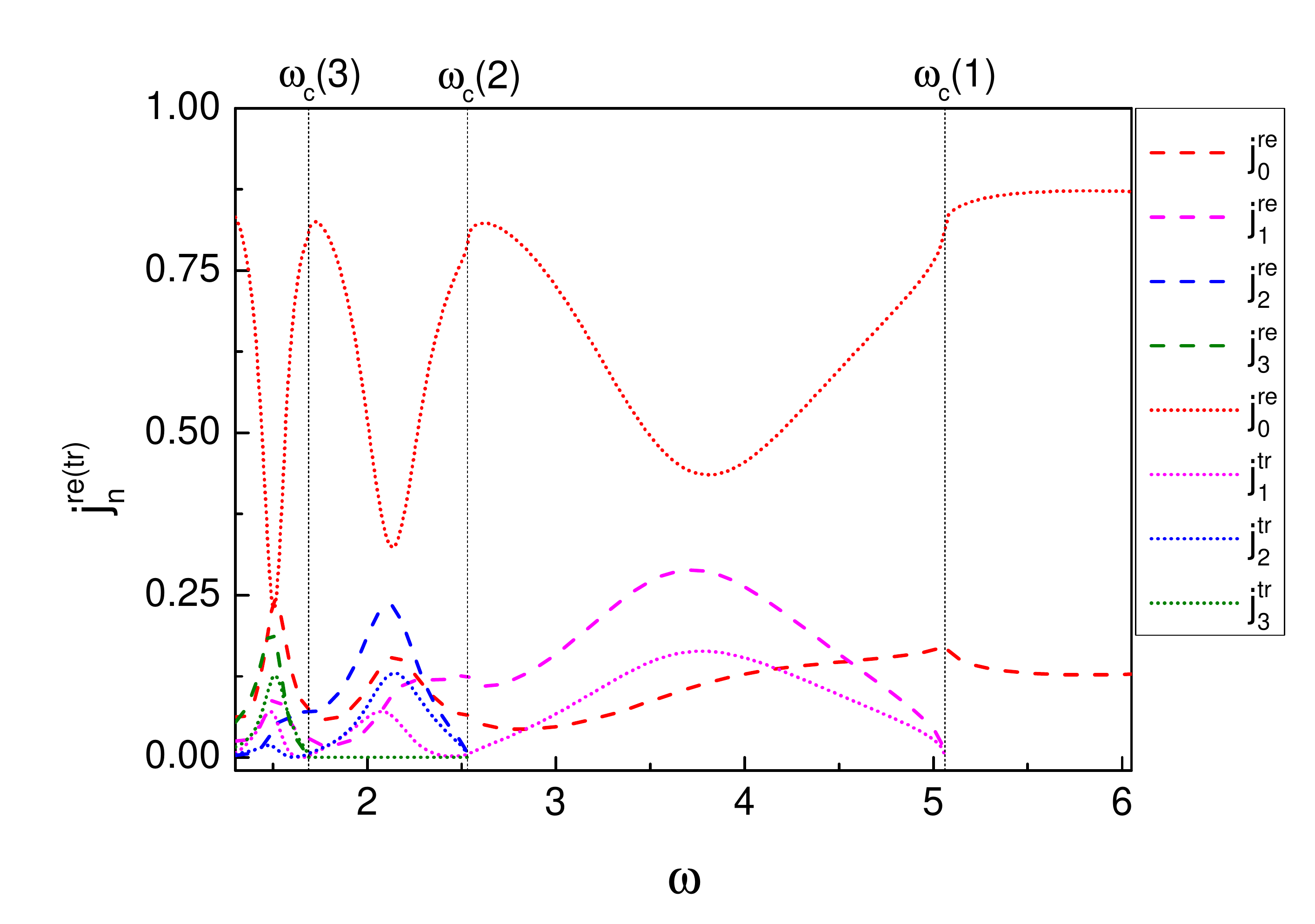}}
  \quad
  \subfloat[][]{\label{omj2} \includegraphics[width=7.5cm]{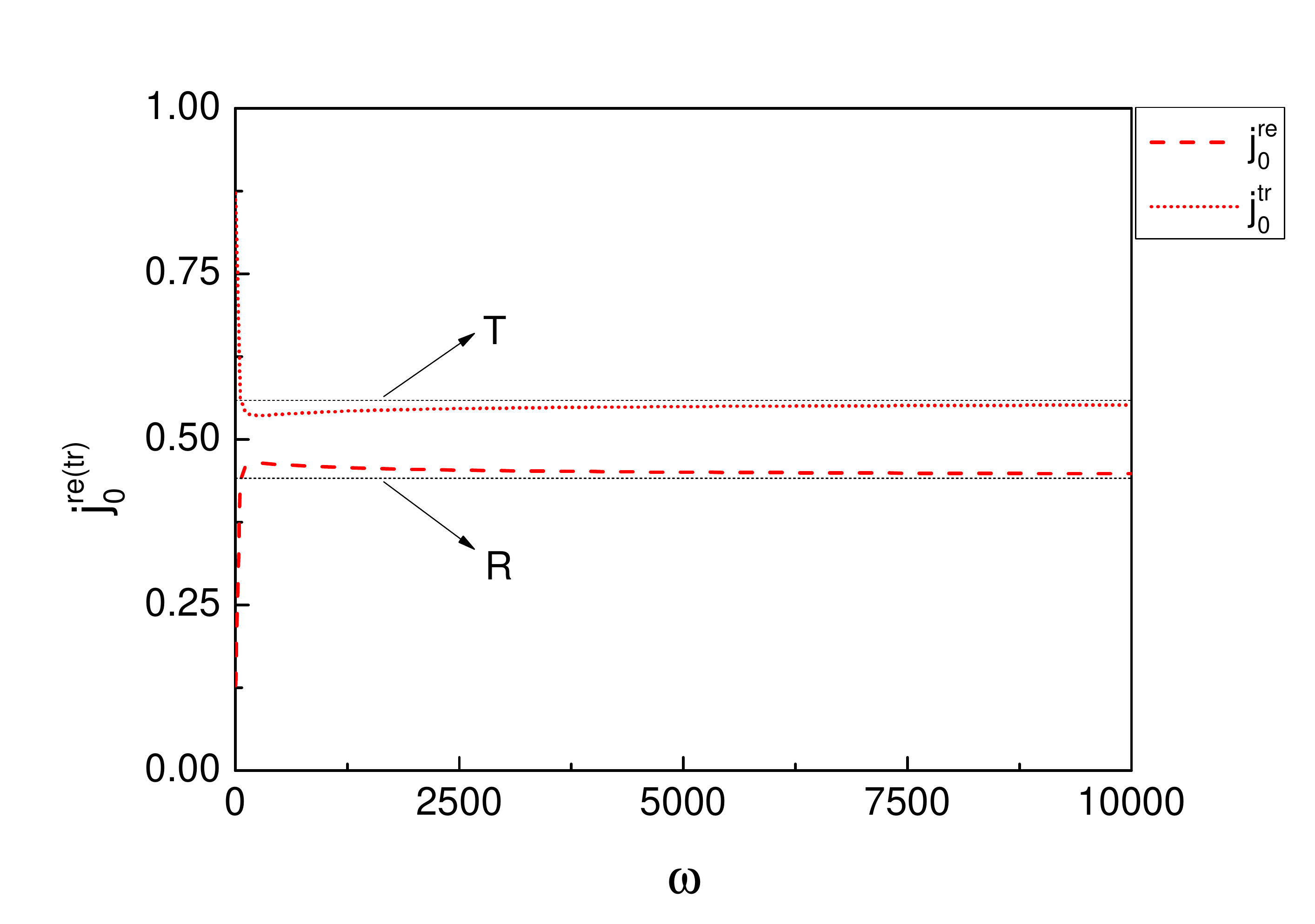}} \\
  \caption{(Color online) The reflection/transmission coefficients
    $j_{n}$ via the coupling strength $\omega$ with parameters $m_{1}=1$,
    $K_{0}=4.5$, $\gamma_{1}=2$, $\gamma_{2}=0$, and (a)
    $\omega\in(1.3,6)$; (b) $\omega\in(6,10^{4})$. The black lines in (b) label the reflection and transmission
    coefficients for a single particle scattered by a delta-function
    potential (see text for details). }~\label{omj0}
\end{figure}

In our model, we assume a harmonic coupling as the binding potential,
and the choice of $\omega$ determines how far the particles of the bound
system can separate from each other.  For a certain $K_{0}$,
the critical coupling stiffness $\omega_{c}(n)$ to excite the $n$-th
internal state is
\begin{equation}
  \omega_{c}(n) = \frac{\hbar K_{0}^{2}}{2(n-l) M}.
  \label{omegac}
\end{equation}
when $\omega\leq\omega_{c}(n)$, the $n$-th mode may be excited. We
exhibit the dependence of $j_{n}$ on the coupling stiffness $\omega$
in Fig.~\ref{omj}. Similar as discussed above, the scattering
potential can only excite the modes whose parity are the same as that
of incident mode for the symmetric
Hamiltonian (Fig.~\ref{omja}, Fig.~\ref{omjb}), while the other modes
can also be excited for the asymmetric Hamiltonian (Fig.~\ref{omjc}).
The greater influence of symmetric Hamiltonian on odd-parity modes
still results in the characteristics "peak" and "valley" for $j_{3}^{re(tr)}$
in Fig.~\ref{omjb}.

We have discussed the situation where both particles interact
individually with the scattering potential above, then we discuss the
case $\gamma_{1}>0$ and $\gamma_{2}=0$ that corresponds to the
situation where particle 2 is only affected indirectly by the
scattering potential via the binding potential. Considering of two
limiting cases, when the coupling stiffness is large enough, the
internal degrees of freedom is confined in the ground state and the
bound system reduces to a single particle, this problem is equivalent
to that of a particle with mass $M$ scattered by a delta-function
potential $\gamma\delta(X)$, the corresponding reflection and
transmission coefficients are
\begin{align}
  R & = \frac{\hbar^{4} K_{0}^{2}}{\hbar^{4} K_{0}^{2} + M^{2}
      \gamma^{2}}, \label{S}\\
  T & = \frac{M^{2} \gamma^{2}}{\hbar^{4} K_{0}^{2} + M^{2}
      \gamma^{2}}. \label{T}
\end{align}
which are marked by the horizontal lines in Fig.~\ref{omj2}. As
Fig.~\ref{omj2} indicates, the reflection and transmission
coefficients $j_{0}^{re},j_{0}^{tr}$ converge to $R$ and $T$
respectively for the extreme large $\omega$.

\begin{figure}[htbp]
  \centering
  \includegraphics[width=7.5cm]{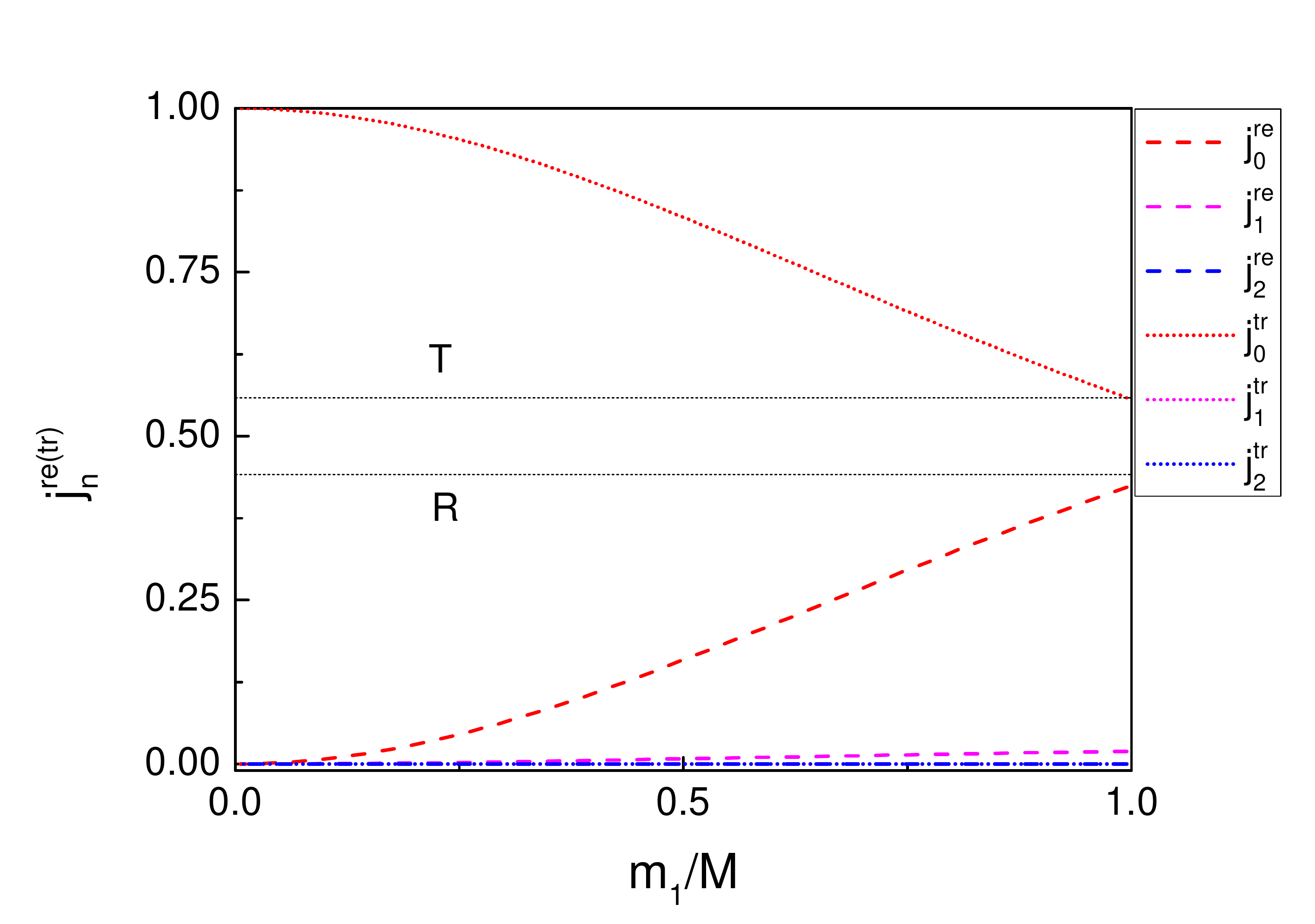}
  \caption{(Color online) The  reflection/transmission coefficients $j_{n}$ via $m_{1}/M$
    with parameters $K_{0}=4.5$, $\omega=2$, $\gamma_{1}=2$, $\gamma_{2}=0$.
    The black lines denote the reflection and transmission
    coefficients for a single particle scattered by a delta-function
    potential (see text for details).}~\label{m1j}
\end{figure}

As we mentioned above, particle 2 is affected by the potential via the
binding potential, the proportion of particle 1 in the bound system
will seriously affect the scattering process. In the limit
$m_{1}\rightarrow 0$, i.e., $m_{1}/M\rightarrow 0$, this problem is
equivalent to that of a particle with mass $M$ passing through the
delta-function potential directly, namely, $j_{0}^{tr}=1,j_{0}^{re}=0$. On the
other extreme, $m_{1}/M\rightarrow 1$, it reduces to the situation
that one particle with $m_{1}$ is scattered by a delta-function
potential, resulting in $j_{0}^{re}=R$ and $j_{0}^{tr}=T$ (see
Fig.~\ref{m1j}).

\begin{figure}[htbp]
  \centering %
  \subfloat[][]{\label{re} \includegraphics[width=7.5cm]{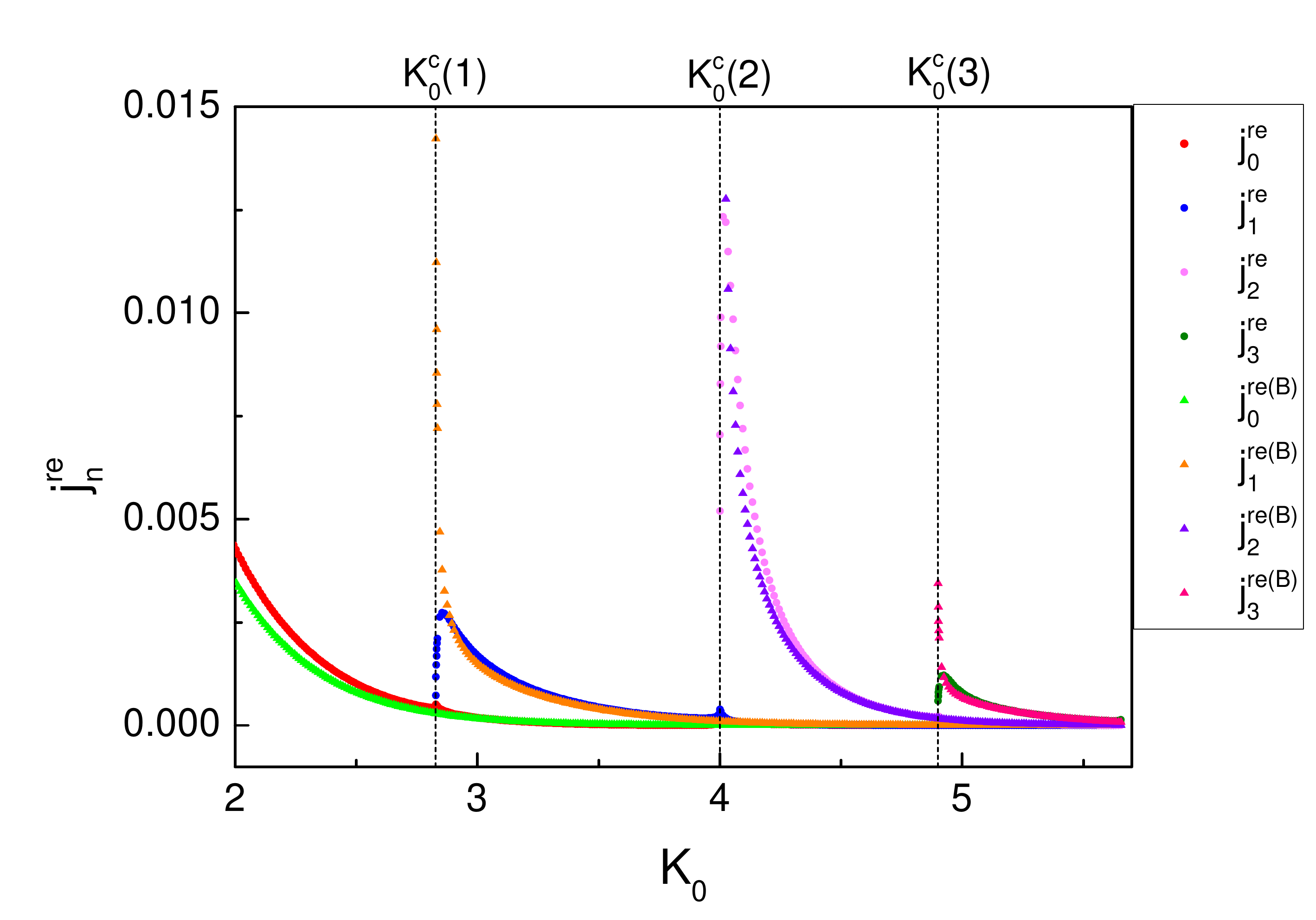}}
  \quad
  \subfloat[][]{\label{tr} \includegraphics[width=7.5cm]{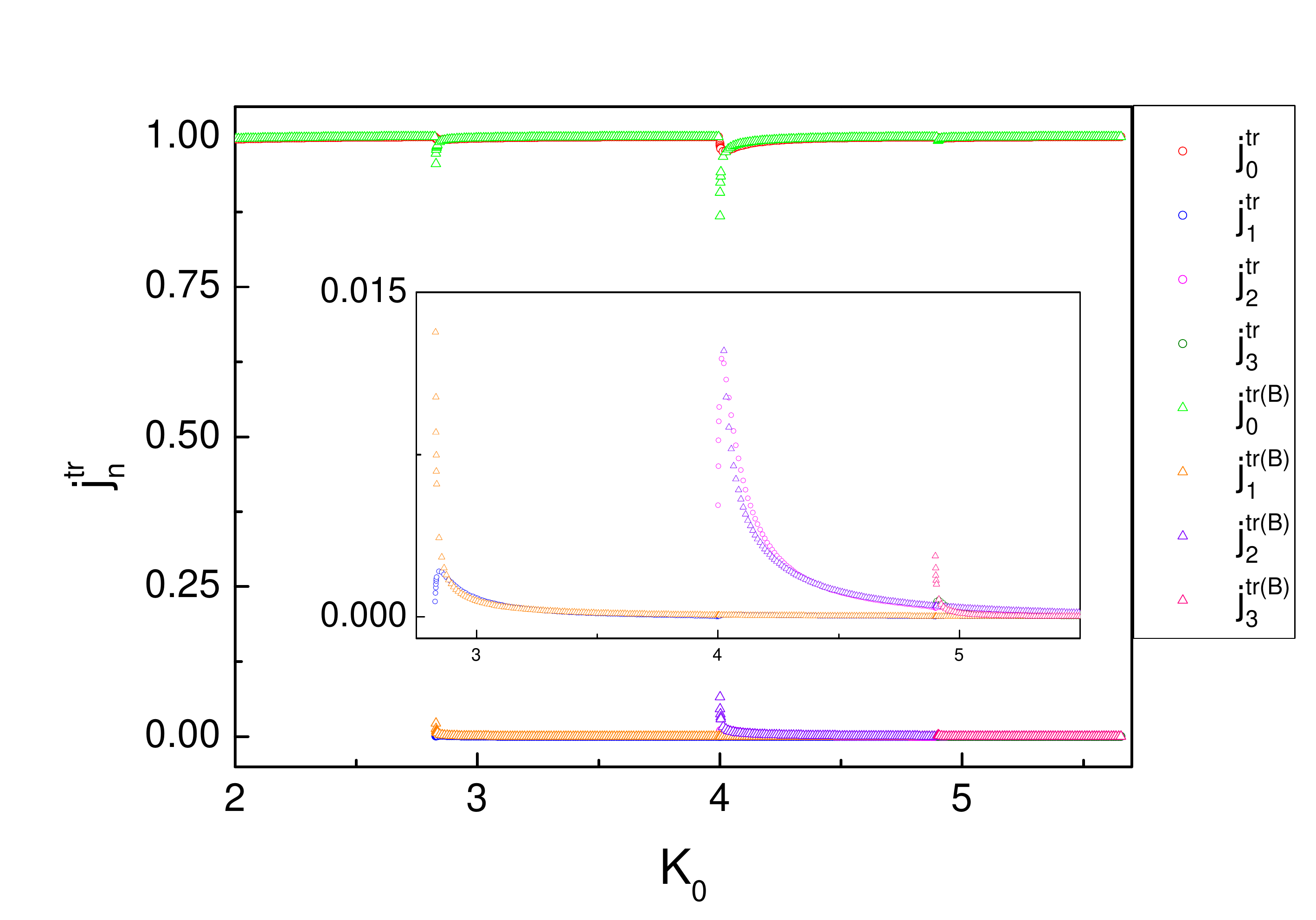}}
  \caption{(Color online) The reflection coefficients $j_{n}^{re}$
    (a) and transmission coefficients $j_{n}^{tr}$ (b) via the potential
    strength $\omega$ with parameters $m_{1}=1.1$, $\omega_{0}=2$,
    $\gamma_{1}=0.1,\gamma_{2}=0.05,l=0$, $j_{n}^{re/tr(B)}$are the
    analytical results within Born approximation.}~\label{kjj}
\end{figure}

In order to test the reliability and accuracy of our results, we
derive the coefficients of reflection and
transmission $j_{n}^{re},j_{n}^{tr}$ by using scattering matrix method
within Born approximation, here $l$ is set to be $l=0$.
\begin{eqnarray}
  j_{0}^{re} & = & |\langle \phi_{-K_0,0}| \mathcal{S}| \phi_{K_0,0}
                   \rangle|^2
        \nonumber\\
  & \approx & \frac{4 \pi^2 M^2}{K_0^2}| \langle \phi_{-K_0,0}|V|
              \phi_{K_0,0} \rangle|^2, \label{R0}
\end{eqnarray}
\begin{align}
  j_{0}^{tr}
  & =  |\langle \phi_{K_0,0}| \mathcal{S}| \phi_{K_0,0} \rangle|^2
    \nonumber\\
  & \approx  1 + \frac{4 \pi^2 M^2}{K_0^2}| \langle \phi_{K_0,0}|V|
    \phi_{K_0,0} \rangle|^2\nonumber\\
  & \quad{} + \frac{4 \pi M}{K_0}\mathbf{Im}{\langle \phi_{K_0,0}|V
    G_0^+V| \phi_{K_0,0}\rangle}
    \nonumber\\
  & =  1 + \frac{4\pi^2 M^2}{K_0^2}| \langle \phi_{K_0,0}|V|
    \phi_{K_0,0} \rangle|^2-\frac{4\pi^2 M^2}{K_0} \sum_{n=0}^{n_{c}}
    \frac{1}{K_{n}}\nonumber\\
  & \quad{} \times \left(| \langle \phi_{-K_n,n}|V| \phi_{K_0,0}
    \rangle|^2+| \langle \phi_{K_n,n}|V| \phi_{K_0,0}
    \rangle|^2\right), \label{T0}
\end{align}
\begin{eqnarray}
  j_{n}^{re}
  & = & | \langle \phi_{-K_n,n}| \mathcal{S}| \phi_{K_0,0} \rangle|^2
        \nonumber\\
  & \approx & \frac{4 \pi^2 M^2}{K_0 K_{n}}| \langle \phi_{-K_n,n}|V|
              \phi_{K_0,0} \rangle|^2 \label{Rn}
\end{eqnarray}
and
\begin{eqnarray}
  j_{n}^{tr}
  & = & | \langle \phi_{K_n,n}| \mathcal{S}| \phi_{K_0,0} \rangle|^2
        \nonumber\\
  & \approx & \frac{4 \pi^2 M^2}{K_0 K_{n}}| \langle \phi_{K_n,n}|V|
              \phi_{K_0,0} \rangle|^2. \label{Tn}
\end{eqnarray}
where $V = \gamma_{1} \delta(X-r_{2}x) +  \gamma_{2} \delta(X+r_{1}x)$ and
\begin{align}
  & \langle \phi_{\pm K_n,n}|V| \phi_{K_0,0} \rangle \nonumber\\
  & =  \frac{ \gamma_{1}}{2 \pi \sqrt{\mu \omega}} \mathbf{e}^{
    \frac{-(K_{n}-K_{0})^2 r_{2}^{2}}{4 \mu \omega}}
    \sqrt{\frac{2^{n}}{n!}} \left(  \frac{i(K_{0}-K_{n})r_{2}}{2
    \sqrt{ \mu \omega}} \right)^{n}
    \nonumber\\
  & \quad{} + \frac{ \gamma_{2}}{2 \pi \sqrt{ \mu \omega}}
    \mathbf{e}^{ \frac{-(K_{n}-K_{0})^2 r_{1}^{2}}{4 \mu \omega}}
    \sqrt{\frac{2^{n}}{n!}} \left(  \frac{i(K_{n}-K_{0})r_{1}}{2
    \sqrt{ \mu \omega}} \right)^{n}.
    \label{Tn}
\end{align}

It's well known that Born approximate is no longer valid for a high
barrier, in Fig.~\ref{kjj} we take $\gamma_{1}=0.1$, $\gamma_{2}=0.05$
and compare our numerical results with analytical ones. For $K_{0}$
away from $K_{0}^{c}(n)$, numerical simulations show qualitative
agreement with these analytical results, while for
$K_{0}\sim K_{0}^{c}(n)$, the analytical results are divergent,
indicating the resonance between the incident mode with the to-be
excited modes.

\section{Discussion and summary}

In summary, we have considered a diatomic bound system to simulate the
composite system, and present how this system is scattered by a
delta-function potential. This could be of importance for a scattering
process of an actual composite system, and even for the testing of the
quantum superposition with a macroscopic object, since we have
considered the internal degrees of freedom. The wave function of the
composite system can be splitted up into two components, leading to
the realization of preparation of a spatial superposition.

When the incoming momentum of the center-of-mass degree of freedom is
large enough, the scattering potential may excite internal states. We
emphasize that the $n>n_{c}$ states could exist in the scattering region.
Physically, these states decay when the system is far away from the
scattering region since the outgoing momentums are imaginary, namely,
there are only $n \leq n_{c}$ internal states being populated at infinite.
Whether the $n \leq n_{c}$ internal states can be excited depends on both the
symmetry of Hamiltonian and the energy condition. All the states under $n_{c}$
could be excited for an asymmetric Hamiltonian, whereas only the states whose parity
are same as incident one could be excited for a symmetric Hamiltonian. The
populations of internal modes are different for reflected and
transmitted components, this should be taken into account in the
experiments of composite system.

We find that the existence of internal degree of freedom can
significantly change the reflection and transmission coefficients of the incident mode
no matter whether the other modes are populated. And the symmetric Hamiltonian has
a more serious impact on the odd-parity internal states than on the even-parity states.

Depending on the coupling strength between the two particles, and also
on the mass of particle 1, the scattering of composite system can
reduce to that of a single particle. Moreover, in the region where
Born approximate is valid, simulation results are in a good accordance
with those of analytical values.

In the present study, we employ a harmonic coupling to mimic the
interaction between particles simply. Our further research within a
general coupling potential, which can describe the process to
transform a diatomic molecular to two atoms, is in process.

\begin{acknowledgments}
  This work is supported by NSF of China (Grant No. 11475254), NKBRSF
  of China (Grant No. 2014CB921202), and The National Key Research and
  Development Program of China (Grant No. 2016YFA0300603).
\end{acknowledgments}

\end{document}